# Heat transport within the Earth

J. Marvin Herndon
Transdyne Corporation
San Diego, CA 92131 USA

Communications: mherndon@san.rr.com

**Abstract:** Numerous attempts have been made to interpret Earth's dynamic processes based upon heat transport concepts derived from ordinary experience. But, ordinary experience can be misleading, especially when underlain by false assumptions. Geodynamic considerations traditionally have embraced three modes of heat transport: conduction, convection, and radiation. Recently, I introduced a fourth, "mantle decompression thermal tsunami" that, I submit, is responsible for emplacing heat at the base of the Earth's crust. Here, I review thermal transport within the Earth and speculate that there might be a fifth mode: "heat channeling", involving heat transport from the core to "hot-spots" such as those that power the Hawaiian Islands and Iceland.

## Introduction

Discovering the true nature of continental displacement, its underlying mechanism, and its energy sources and modes of heat transport are among the most fundamental geoscience challenges. The seeming continuity of geological structures and fossil life-forms on either side of the Atlantic Ocean and the apparent "fit' of their opposing coastlines led Snider-Pellegrini [1] to propose in 1858, as shown in Figure 1, that the Americas were at one time connected to Europe and Africa and subsequently separated, opening the Atlantic Ocean. Half a century later, Wegener promulgated a similar concept, with more detailed justification, that became known as "continental drift" [2]. According to Wegener's theory, in the past the continents were united, but 300 million years ago broke apart with the pieces drifting through the ocean floor to their present locations.

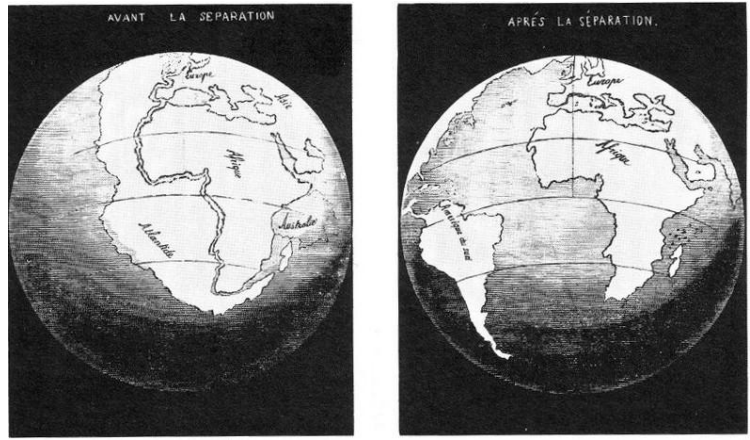

**Figure 1.** The opening of the Atlantic Ocean, reproduced from (Snider-Pellegrini, 1858) [1].

Any theory of continental displacement requires a physically realistic mechanism and an adequate energy source. In 1931, Holmes elaborated upon Bull's [3] concept of mantle convection, originally suggested to explain mountain building, and proposed it as a mechanism for continental drift, publishing the illustration reproduced as Figure 2 [4].

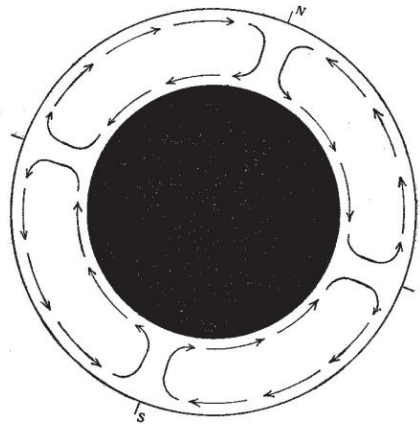

**Figure 2.** Schematic representation of mantle convection, from [4]. Reproduced with permission of the Geological Society of Glasgow.

Three decades later the discovery of ocean-floor magnetic striations -- symmetric to the mid-ocean ridge and progressively older with distance from it -- were well explained by "seafloor spreading", which became a crucial component of plate tectonics. The idea that seafloor is extruded from the mid-ocean ridges, moves across the ocean basin and is "subducted" into submarine trenches reinforced and seemed to justify the concept of mantle convection, as illustrated by the U. S. Geological Survey diagram reproduced as Figure 3. To many, the explanation seemed so correct that mantle convection "must" exist. But there are serious problems with the continental-drift/plate-tectonics hypothesis and with the concept of mantle convection.

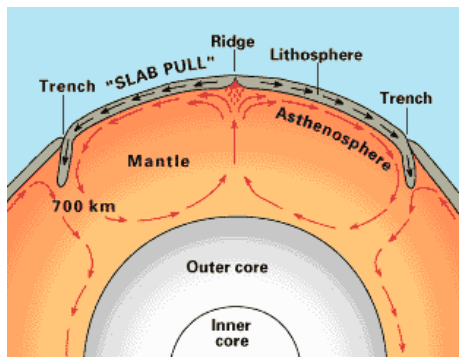

**Figure 3.** U. S. Geological Survey schematic representation of mantle convection associated with plate tectonics theory.

Seventy years ago, Elsasser [5] published his idea, still popular today, that the geomagnetic field is produced by convective motions in the Earth's fluid, electrically-conducting core, interacting with Coriolis forces produced by planetary rotation, creating a dynamo mechanism, a magnetic amplifier. Although the geomagnetic field reverses polarity irregularly, it has been remarkably stable for long periods of time, including intervals as long as 40 million years without reversals. Elsasser's convection-driven dynamo mechanism seemed to explain so well the generation of the geomagnetic field that for decades geophysicists believed convection in the Earth's fluid core "must" exist. But there are serious problems with the concept of Earth-core convection.

The considerable confusion in the scientific literature as to the nature of Earth's dynamics and internal heat sources that can be traced to two erroneous assumptions: (1) Since 1940, that Earth's chemical and mineral composition resembles an ordinary chondrite meteorite; and, (2) Since 1963, that Earth formed from dust, condensed from an atmosphere of solar composition at very low pressure, $10^{-5}$ bar, that gathered into progressively larger grains, then rocks, then planetesimals, finally planet Earth. Here I review the energy sources and modes of heat transport within the Earth.

## Earth's Internal Heat Sources

Heat from radioactive-decay of $^{235}U$, $^{238}U$, $^{232}Th$, and $^{40}K$ has long been (wrongly) considered as the main energy source for geodynamics processes, geomagnetic field generation, and for the Earth's heat loss. For more than half a century, geophysicists have made measurements of near-surface continental and oceanic heat flow with the aim of determining the Earth's heat loss. Pollack *et al.* [6] estimate a global heat loss of 44.2 terawatts (TW, 1 TW=$10^{12}$ W) based upon 24,774 observations at 20,201 sites. The problem is that radioactive-decay heat alone cannot satisfy just the global heat loss requirements. Estimates of present-day global radiogenic heat production, based upon chondritic abundances of $^{235}U$, $^{238}U$, $^{232}Th$, and $^{40}K$, typically range from 19 TW to 31 TW. These represent upper limits through the tacit, unrealistic assumption of very rapid heat transport irrespective of radionuclide locations [7]. Moreover, it has long been known geomagnetic field originates at or near the center of the Earth [8] so there must be an energy source there.

Confusion as to the nature of location and nature of radionuclide energy sources within the Earth stems from the mistaken belief, prevalent for the past seventy years, that the Earth is like an ordinary chondrite meteorite rather than, as I discovered thirty years ago, a highly-reduced enstatite chondrite [9-13]. In ordinary chondrites, which formed under more oxidizing conditions than enstatite chondrites [14], all of the radionuclides are found in the silicate portion. It has been (wrongly) assumed therefore that these would occur exclusively in the Earth's mantle and crust. Reports, however, have suggested that at high pressures $^{40}K$ might occur in the Earth's core [15]. The absence of core-heat sources in an "ordinary-chondritic Earth" led to the *ad hoc* suggestion, without corroborating evidence, that the inner core is growing by freezing, releasing the heat of crystallization which hypothetically provides useful energy rather than just slowing the assumed rate of freezing.

The identification of the endo-Earth (lower mantle plus core) with an enstatite chondrite [13] made it possible for me to deduce that the bulk of Earth's uranium resides within the core and to demonstrate the feasibility of its functioning as a natural nuclear fission reactor [16-20]. The nuclear georeactor is an unanticipated deep-Earth energy source that, I submit, produces the Earth's magnetic field [16-22]. Energy production from the nuclear fission of uranium is significantly greater than from its radioactive decay, but may consume uranium at a much greater rate. It is an open question as to whether thorium, possibly also in the Earth's core, exists under circumstances that might permit it to be converted to fissionable $^{234}$U and thereby produce more energy than by radioactive decay alone.

The identification of the endo-Earth with an enstatite chondrite made it possible for me to deduce from thermodynamic considerations the circumstances of Earth's early formation as a Jupiter-like gas-giant and to reveal another major, unanticipated energy source, the stored heat of protoplanetary compression [23]. This vast energy source, I submit, is responsible for fracturing Hadean Earth's 100% closed, contiguous, continental-rock shell, for decompressing Earth and for powering Earth's compression-driven geology, as described by whole-Earth decompression dynamics [24, 25], and for emplacing heat at the base of the crust [26].

### Heat Emplacement at the Base of the Crust

Since 1939, scientists have been measuring the heat flowing out of continental-rock [27, 28] and, since 1952, heat flowing out of ocean floor basalt [29]. Continental-rock contains much more of the long-lived radioactive nuclides than does ocean floor basalt. So, when the first heat flow measurements were reported on continental-rock, the heat was naturally assumed to arise from radioactive decay. But later, ocean floor heat flow measurements showed more heat flowing out of the ocean floor basalt than out of continental-rock. This seemingly paradoxical result, I suggest, arises from a previously unanticipated mode of heat transport that emplaces heat at the base of the crust, which I call mantle decompression thermal tsunami [26].

As the Earth decompresses, heat must be supplied to replace the lost heat of protoplanetary compression. Otherwise, decompression would lower the temperature, which would impede the decompression process. Heat generated deep within the Earth may enhance mantle decompression by replacing the lost heat of protoplanetary compression. The resulting decompression, beginning within the mantle, will tend to propagate throughout the mantle, like a tsunami, until it reaches the impediment posed by the base of the crust. There, crustal rigidity opposes continued decompression, pressure builds and compresses matter at the mantle-crust-interface, resulting in compression heating. Ultimately, pressure is released at the surface through volcanism and through secondary decompression crack formation and/or enlargement. Mantle decompression thermal-tsunami poses a new explanation for heat emplacement at the base of the crust, which may be involved in earthquakes and volcanism, as these geodynamic processes appear concentrated along secondary decompression cracks, and may be involved in the formation of abiotic hydrocarbons [25, 30].

## Heat from the Earth's Core

Helium, trapped in volcanic lava, is observed is a variety of geological settings. The $^3$He/$^4$He ratios measured in basalt extruded at the mid-ocean ridges are remarkably constant, averaging 8.6 times the same ratio measured in air. The $^3$He/$^4$He ratios measured in lava from 18 "hot-spots" around the globe, such as the Hawaiian Islands, are greater than 10 times the value in air. Georeactor numerical simulations [19, 20] demonstrate fission-product helium in the range of compositions observed in basalt, and indicate a progressive rise in $^3$He/$^4$He ratios over time as uranium fuel is consumed by nuclear fission and radioactive decay. The high helium ratios measured in hot-spot lavas appear to be the signature of georeactor-produced heat and helium. In certain instances, thermal structures beneath hot-spots, sometimes called mantle plumes, as imaged by seismic tomography [31, 32] extend to the interface of the core and lower mantle, further reinforcing their georeactor-heat origin.

The Hawaiian Islands and Iceland are two high $^3$He/$^4$He, ocean-floor-piercing, currently erupting hot-spots with seismic imaging indicating that their heat sources arise from the core-mantle boundary. Mjelde and Faleide [33] recently discovered a periodicity and synchronicity through the Cenozoic in lava outpourings from these two hot-spots that Mjelde et al. [34] suggest may arise from variable georeactor heat-production.

As well as piercing the ocean crust, high $^3$He/$^4$He hot-spot volcanism presently occurs beneath continental masses: Yellowstone (U. S. A.) and Afar in the East African Rift System being two current examples. The massive flood basalts of the Deccan Traps of India (65 million years ago) [35] and the Siberian Traps (250 million years ago) [36] are likewise characterized by high $^3$He/$^4$He ratios indicating georeactor-heat origin.

Tomographic images of so-called mantle plumes beneath hot-spots have become increasingly important for geological understanding. But, even with the advent of seismic tomography, there is still considerable controversy as to the true nature of mantle plumes and to the question of whether or not mantle plumes actually exist [37, 38]. The mantle plume concept had its origins in Wilson's 1963 suggestion [39] that the volcanic arc comprised of the Hawaiian Islands formed as seafloor moved across a persistent, fixed "hot-spot". In 1971, Morgan [40] proposed that hot-spots are manifestations of convection in the lower mantle. Here I describe the reasons that mantle convection is physically impossible, and speculate on the idea of "heat channeling" as a means of heat transport from the core-mantle boundary to the surface.

**Mantle Heat Channeling**

Since the 1930's, convection has been assumed to occur within the Earth's mantle [4] and, since the 1960's has been incorporated as an absolutely crucial component of seafloor spreading in plate tectonics theory. Instead of looking questioningly at the process of convection, many have assumed without corroborating evidence that mantle convection "must" exist.

Chandrasekhar [41] described convection in the following way: "The simplest example of thermally induced convection arises when a horizontal layer of fluid is heated from below and an adverse temperature gradient is maintained. The adjective 'adverse' is used to qualify the prevailing temperature gradient, since, on account of thermal expansion, the fluid at the bottom becomes lighter than the fluid at the top; and this is a top-heavy arrangement which is potentially unstable. Under these circumstances the fluid will try to redistribute itself to redress this weakness in its arrangement. This is how thermal convection originates: It represents the efforts of the fluid to restore to itself some degree of stability."

The lava lamp, invented by Smith [42], affords an easy-to-understand demonstration of convection at the Earth's surface. Heat warms a blob of wax at the bottom, making it less dense than the surrounding fluid, so the blob floats to the surface, where it loses heat, becomes denser than the surrounding fluid and sinks to the bottom. Convection is applicable in circumstances wherein density is constant except as altered by thermal expansion; in the lava-lamp, for example, but not in the Earth's mantle. The Earth's mantle is "bottom-heavy", i.e., its density at the bottom is about 62% greater than its top (Figure 4). The potential decrease in density by thermal expansion, <1%, cannot make the mantle "top-heavy" as described by Chandrasekhar. Thus mantle convection cannot be expected to occur.

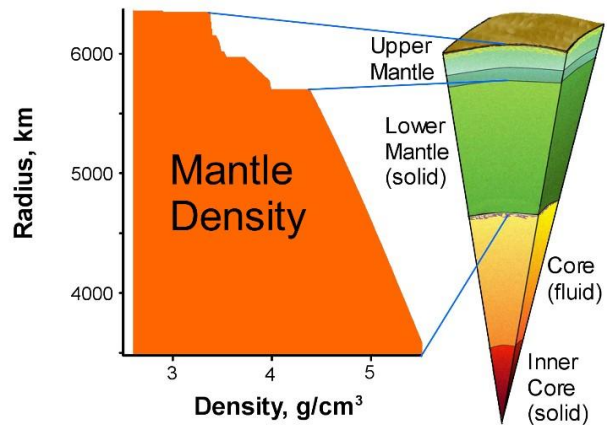

**Figure 4.** Density as a function of radius in the Earth's mantle [44]. The representation of the Earth's major parts shows (unlabeled) the core-floaters, CaS and MgS, at the top of the core that, I submit, are responsible for the seismic "roughness" at D″, the core-mantle boundary. The georeactor at the center of Earth, one ten-millionth the mass of the fluid core is not shown.

Mantle convection is often (wrongly) asserted to exist on the basis of a high calculated Rayleigh Number [43], which was derived to quantify the onset of instability in a thin, horizontal layer of *incompressible* fluid of *uniform density*, except as altered by thermal expansion when heated from beneath. The Rayleigh Number is not applicable to the Earth's mantle, which is neither incompressible nor of uniform density.

It is instructive to apply the principle upon which submarines operate, "neutral buoyancy", to the Earth's mantle. The idea is that a heated "parcel" of bottom mantle matter, under the physically-unrealistic assumption of ideal, optimum conditions, will float upward to come to rest at its "neutral buoyancy", the point at which its own density is the same as the prevailing mantle density.

Consider a "parcel" of matter at the base of the Earth's lower mantle existing at the prevailing temperature, $T_0$, and having density, $\rho_0$, indicated by the data upon which Figure 4 is based [44]. Now, suppose that the "parcel" of bottom-mantle matter is selectively heated to temperature ΔT degrees above $T_0$. The "parcel" will expand to a new density, $ρ_z$, given by

$$ρ_z = ρ_0 (1-αΔT)$$

where α is the volume coefficient of thermal expansion at the prevailing temperature and pressure.

Now, consider the resulting dynamics of the newly expanded "parcel". Under the assumption of ideal, optimum conditions, the "parcel" will suffer no heat loss and will encounter no resistance as it floats upward to come to rest at its "neutral buoyancy", the point at which its own density is the same as the prevailing mantle density. The Earth-radius of the "neutral buoyancy" point thus determined can be obtained from the data upon which Figure 4 is based; the "maximum float distance" simply is the difference between that value and the Earth-radius at the bottom of the lower mantle.

The relationship between "maximum float distance" and ΔT thus calculated for the lower mantle is shown in Figure 5. At the highest ΔT shown, the "maximum float distance" to the point of "neutral buoyancy" is <25 km, just a tiny portion of the 2230 km distance required for lower mantle convection, and nearly 2900 km required for whole-mantle convection. Even with the assumed "ideal, optimum conditions" and an unrealistically great ΔT = 600°K, an error in the value of α by two orders of magnitude would still not cause the "maximum float distance" to reach 2900 km. I use "ideal" for purposes of illustration, but in nature "ideal" does not exist, and only in certain quite limited instances is ideal behavior even approached.

Decades of belief that mantle convection "must" exist has resulted in a plethora of mantle convection models that, of course, purport to show that mantle convection is possible under certain assumed conditions. Generally, models begin with a preconceived result that is invariably achieved through result-selected assumptions. Although rarely, if ever, stated explicitly, in convection models, the mantle is tacitly assumed to behave as an ideal gas.

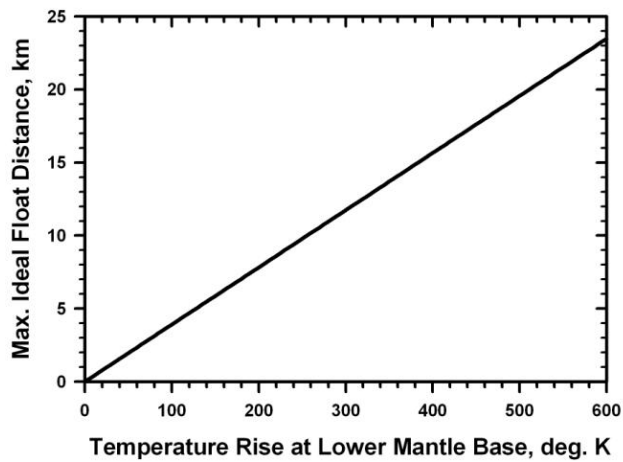

**Figure 5.** The "maximum float distance" to "neutral buoyancy" from the base of the lower mantle as a function of "parcel" temperature rise. The value used for the coefficient of thermal expansion, $\alpha=0.37 \times 10^{-5}$ K$^{-1}$, is from the standard reference state value of MgSiO$_3$ perovskite [48], reduced by 80% to take into account lower mantle base temperature and pressure, according to [49].

Stellar convection models involved a gravitationally compressed system of H$_2$ and He gas at ~5000K that is thought to approach ideal gas behavior, i.e., no viscosity, hence, no viscous loss. In those models a heated parcel of ideal-gas expands and rises, never losing heat to its surroundings, and never coming to rest at "neutral buoyancy". The parcel maintains pressure equilibrium with its surroundings as it begins to rise, decompressing and expanding against progressively lower pressure, while maintaining its initial heat perturbation. The only impediment to such ideal-gas convection is if heat can be transported more rapidly by conduction and/or radiation than by convection.

Mantle convection models typically apply the same reasoning and assumptions as stellar convection models. A heated parcel of mantle matter is assumed to float ever-upward decreasing in density, never reaching "neutral buoyancy", while maintaining its heat content. But the mantle is not an ideal gas; it is a crystalline solid, not even a super-cooled liquid like glass. But, like its stellar counterpart, it assumed to behave "adiabatically", i.e., to maintain the parcel's initial heat perturbation, suffering no heat loss, even although in reality the mantle: (1) is extremely viscous and thus subject to viscous losses; (2) potentially moves by convection at a rate not too different from the rate heat is conducted; (3) has compositionally-different layers; (4) may have crystalline phase boundaries; and (5) possesses unknown rheological properties. Earthquakes, for example, occur within the mantle to depths of about 690km and signal the catastrophic release of pent-up stress. Processes and properties such as these, I submit, would readily block mantle convection. And, since whole-Earth decompression dynamics, mantle convection is not necessary to explain the observed seafloor topography better than by plate tectonics [45]. The underlying principles of mantle convection, however, might operate on a micro-scale and contribute in a yet undetermined way to a process of mantle heat channeling.

Envision heat originating at a point on the Earth's core-mantle boundary. If thermal conduction alone were involved in its transport, one might expect the heat to be conducted to regions of lower temperature in a more-or-less hemispherical pattern. But seismic tomography appears to image vertical, column-like heat paths, for example, beneath the Hawaiian Islands, that cannot represent-matter transport by convection for the reasons described above.

Water, uniformly distributed upon soil, often peculates downward by gravity in a non-uniform way, forming channels through paths of less resistance. An analogous process might occur in the Earth's mantle for the upward-channeling of heat. Innumerable-layers of buoyancy-driven micro-convection in conjunction with conduction, I speculate, operates to directionally-bias and/or augment the flow of core-derived heat upward.

### Heat Transport within the Earth's Core

As is the case for the Earth's mantle, justification for Earth-core convection cannot be obtained by calculating the Rayleigh Number because the Earth's core is neither incompressible nor of uniform density. Although the Earth's core is liquid, it is "bottom heavy", i.e., its density at the bottom is about 23% greater than its top. The potential decrease in density by thermal expansion, <1%, cannot make the core "top-heavy" as described by Chandrasekhar; thus convection is not to be expected. But, there is an even more serious impediment to Earth-core convection.

For sustained convection to occur, heat brought from the core-bottom must be efficiently removed from the core-top to maintain the "adverse temperature gradient" described by Chandrasekhar, i.e., the bottom being hotter than the top. But, efficient heat removal is physically impossible because the Earth's core is wrapped in an insulating silicate blanket, the mantle, 2900 km thick that has significantly lower thermal conductivity, lower heat capacity, and greater viscosity than the Earth's core. Heat transport within the Earth's fluid core must therefore occur mainly by thermal conduction, not convection.

The geomagnetic implication is quite clear: Either the geomagnetic field is generated by a process other than the convection-driven dynamo-mechanism, or there exists another fluid region within the deep-interior of Earth which can sustain convection for extended periods of time. I have provided the reasonable basis to expect long-term stable convection in the georeactor sub-shell, and proposed that the geomagnetic field is generated therein by the convection-driven dynamo mechanism [21, 22]. Heat produced by the georeactor's nuclear sub-core causes convection in the surrounding fluid radioactive-waste sub-shell; heat is removed from the top of the sub-shell by a massive, thermally-conducting heat-sink (the inner core) that is surrounded by an even more massive, thermally-conducting heat-sink (the fluid core).

There are fundamental differences in convection-driven dynamo action in the georeactor sub-shell than in the Earth's core, as has long been wrongly believed: (1) The georeactor sub-shell contains a substantial quantity of continuously-supplied, neutron-rich, radioactive fission products that beta decay, producing electrons which can generate magnetic seed-fields for amplification; (2) The dimensions, mass, and inertia are orders of magnitude less than those of the Earth's core, meaning that changes in the geomagnetic field, including reversals and excursions, can take place on much shorter time-scales than previously thought, in accord with observations [46]; and (3) External effects may assume greater importance, for example, super-intense bursts of solar wind might induce electrical currents and consequently ohmic heating in the georeactor sub-shell, perhaps destabilizing convection and leading to magnetic reversals.

**Scholarium**

Science is very much a logical progression of understanding through time. Advances are frequently underpinned by ideas and understandings developed in the past, sometimes under circumstances which may no longer hold the same degree of validity [47]. All too often, scientists, being distinctly human creatures of habit, plod optimistically along through time, eagerly looking toward the future, but rarely looking with question at circumstances from the past which have set them upon their present courses. Instead of making models based upon assumptions, one might look questioningly at past developments, and ask whether these are in conflict with the properties of matter as now known. Correcting past faltering leads to future progress.

## References

1.  Snider-Pellegrini, A., *La Création et ses mystères dévoilés (Creation and its Mysteries Unveiled)*. 1858, Paris.

2.  Wegener, A. L., Die Entstehung der Kontinente. *Geol. Rundschau*, 1912, **3,** 276-292.

3.  Bull, A.J., A hypothesis of mountain building. *Geol. Mag.*, 1921, **58**, 364-397.

4.  Holmes, A., Radioactivity and Earth movements. *Trans. geol. Soc. Glasgow* 1928-1929, 1931, **18**, 559-606.

5.  Elsasser, W. M., On the origin of the Earth's magnetic field. *Phys. Rev.*, 1939, **55**, 489-498.

6.  Pollack, H. N., Hurter, S. J. and Johnson, J. R., Heat flow from the Earth's interior: Analysis of the global data set. *Rev. Geophys.*, 1993, **31**(3), 267-280.

7.  Kellogg, L.H., Hager, B. H. and van der Hilst, R. D., Compositional stratification in the deep mantle. *Science*, 1999, **283**, 1881-1884.

8.  Gauss, J. C. F., *Allgemeine Theorie des Erdmagnetismus: Resultate aus den Beobachtungen des magnetischen Vereins in Jahre 1838*. 1838, Leipzig. 73.

9. Herndon, J. M., The nickel silicide inner core of the Earth, *Proc. R. Soc. Lond.*, **A368**, 1979, 495-500.

10. Herndon, J. M., The chemical composition of the interior shells of the Earth. *Proc. R. Soc. Lond.*, 1980, **A372**, 149-154.

11. Herndon, J. M., The object at the centre of the Earth. *Naturwissenschaften*, 1982, **69**, 34-37.

12. Herndon, J. M., Composition of the deep interior of the earth: divergent geophysical development with fundamentally different geophysical implications, *Phys. Earth Plan. Inter.*, **105**, 1998, 1-4.

13. Herndon, J. M., Scientific basis of knowledge on Earth's composition", *Curr. Sci.*, **88(7)**, 2005, 1034-1037.

14. Herndon, J. M., Discovery of fundamental mass ratio relationships of whole-rock chondritic major elements: Implications on ordinary chondrite formation and on planet Mercury's composition, *Curr. Sci.*, **93(3)**, 2007, 394-399.

15. Murthy, V. R., van Westernen, W.,and Fei, Y., Experimental evidence that potassium is a substantial radioactive heat source in planetary cores. *Nature*, 2003, **423**, 163-165.

16. Herndon, J. M., Feasibility of a nuclear fission reactor at the center of the Earth as the energy source for the geomagnetic field. *J. Geomag. Geoelectr.*, 1993, **45**, 423-437.

17. Herndon, J. M., Planetary and protostellar nuclear fission: Implications for planetary change, stellar ignition and dark matter. *Proc. R. Soc. Lond.*, 1994, **A455**, 453-461.

18. Herndon, J. M., Sub-structure of the inner core of the earth. *Proc. Nat. Acad. Sci. USA*, 1996, **93,** 646-648.

19. Herndon, J. M., Nuclear georeactor origin of oceanic basalt $^3$He/$^4$He, evidence, and implications. *Proc. Nat. Acad. Sci. USA*, 2003, **100**(6), 3047-3050.

20. Hollenbach, D. F. and Herndon J. M., Deep-earth reactor: nuclear fission, helium, and the geomagnetic field. *Proc. Nat. Acad. Sci. USA*, 2001, **98**(20), 11085-11090.

21. Herndon, J. M., Nuclear georeactor generation of the earth's geomagnetic field. *Curr. Sci.*, 2007, **93**(11), 1485-1487.

22. Herndon, J. M., Nature of planetary matter and magnetic field generation in the solar system. *Curr. Sci.*, 2009, **96,** 1033-1039.

23. Herndon, J. M., Solar System processes underlying planetary formation, geodynamics, and the georeactor, *Earth, Moon, and Planets*, **99(1)**, 2006, 53-99.

24. Herndon, J. M., Whole-Earth decompression dynamics. *Curr. Sci.*, 2005, **89**(10), 1937-1941.

25. Herndon, J. M., Impact of recent discoveries on petroleum and natural gas exploration: Emphasis on India. *Curr. Sci.*, 2010, **98,** 772-779.

26. Herndon, J. M., Energy for geodynamics: Mantle decompression thermal tsunami. *Curr. Sci.*, 2006, **90**, 1605-1606.

27. Benfield, A. F., Terrestrial heat flow in Great Britain. *Proc. R. Soc. Lond.*, 1939, **A 173**, 428-450.

28. Bullard, E. C., Heat flow in South Africa. *Proc. R. Soc. Lond.*, 1939, **A 173**, 474-502.

29. Revelle, R. and Maxwell, A. E., Heat flow through the floor of the eastern North Pacific Ocean. *Nature*, 1952, **170**, 199-200.

30. Herndon, J. M., Enhanced prognosis for abiotic natural gas and petroleum resources. *Curr. Sci*., 2006, **91**(5), 596-598.

31. Bijwaard, H. and Spakman W., Tomographic evidence for a narrow whole mantle plume below Iceland. *Earth Planet. Sci. Lett*., 1999, **166**, 121-126.

32. Nataf, H. -C., Seismic Imaging of Mantle Plumes. *Ann. Rev. Earth Planet. Sci*., 2000, **28**, 391-417.

33. Mjelde, R. and Faleide J. I., Variation of Icelandic and Hawaiian magmatism: evidence for co-pulsation of mantle plumes? *Mar. Geophys. Res*., 2009, **30**, 61-72.

34. Mjelde, R., Wessel, P. and Müller, D., Global pulsations of intraplate magmatism through the Cenozoic. *Lithosphere*, 2010, **2**(5), 361-376.

35. Basu, A. R., et al., Early and late alkali igneous pulses and a high-$^3$He plume origin for the Deccan flood basalts. *Sci*., 1993, **261**, 902-906.

36. Basu, A. R., et al., High-$^3$He plume origin and temporal-spacial evolution of the Siberian flood basalts. *Sci*., 1995, **269**, 882-825.

37. Foulger, G. R., et al., The great plume debate. *Eos, Trans. Am. Geophys. U*., 2006, **87**(7), 76, 80.

38. Sankaran, A. V., The row over earth's mantle plume concept. *Curr. Sci*., 2004, **87**(9), 1170-1172.

39. Wilson, J. T., A possible origin of the Hawaiian Islands. *Can. J. Phys*., 1963, **41** 863-870.

40. Morgan, W. J., Convection plumes in the lower mantle. *Nature*, 1971, **230**, 42-43.

41. Chandrasekhar, S., Thermal Convection. *Proc. Amer. Acad. Arts Sci.*, 1957, **86**(4), 323-339.

42. Smith, D.G., *Display Devices*. U. S. Patent 3,387,396. 1968.

43. Lord_Rayleigh, On convection currents in a horizontal layer of fluid where the higher temperature is on the under side. *Phil. Mag.*, 1916, **32**, 529-546.

44. Dziewonski, A. M. and Anderson D. A., Preliminary reference Earth model. *Phys. Earth Planet. Inter.*, 1981, **25**, 297-356.

45. Herndon, J. M., Energetic to the core: The whole Earth story. *American Scientist*, 2010, in press.

46. Coe, R. S. and Prevot, M., Evidence suggesting extremely rapid field variation during a geomagnetic reversal. *Earth Planet. Sci. Lett.*, 1989, **92**, 192-198.

47. Herndon, J. M., Inseparability of science history and discovery. *Hist. Geo Space Sci.*, 2010, **1**, 25-41.

48. Oganov, A. R., Brodholt, J. P. and Price, G. D., Ab initio elasticity and thermal equation of state of MgSiO3 perovskite. *Earth Planet. Sci. Lett.*, 2001, **184**, 555-560.

49. Birch, F., Elasticity and constitution of the Earth's interior. *J. Geophys. Res.*, 1952, **57**(2), 227-286.